\documentclass[aps,prb,twocolumn,showpacs,floatfix,longbibliography]{revtex4-2}
\usepackage{graphicx}% Include figure files
\usepackage{dcolumn}% Align table columns on decimal point
\usepackage{bm}% bold math
\usepackage{multirow}
\usepackage{float}
\usepackage[colorlinks,citecolor=blue,linkcolor=red,urlcolor=blue]{hyperref}

\begin{document}
%\begin{CJK*}{GBK}{}
\title{Cascade of the delocalization transition in a non-Hermitian interpolating Aubry-Andr{\'e}-Fibonacci chain}
\author{Liang-Jun Zhai$^{1,2}$}
\author{Guang-Yao Huang$^{3}$}
\email{guangyaohuang@quanta.org.cn}
\author{{Shuai Yin$^{4}$ }}
\email{yinsh6@mail.sysu.edu.cn}
\affiliation{$^1$The school of mathematics and physics, Jiangsu University of Technology, Changzhou 213001, China}
\affiliation{$^2$Department of Physics, Nanjing University, Nanjing 210093, China}
\affiliation{$^3$Institute for Quantum Information $\&$  State Key Laboratory of High Performance Computing, College of Computer, National University of Defense Technology, Changsha 410073, China}
\affiliation{$^4$School of Physics, Sun Yat-Sen University, Guangzhou 510275, China}
%\date{\today}
\begin{abstract}
In this paper, the interplay of the non-Herimiticity and the cascade of delocalization transition in the quasi-periodic chain is studied.
The study is applied in a non-Hermitian interpolating Aubry-Andr{\'e}-Fibonacci (IAAF) model, which combines the non-Hermitian Aubry-Andr{\'e} (AA)
model and the non-Hermitian Fibonacci model through a varying parameter, and the non-Hermiticity in this model is introduced by the non-reciprocal hopping.
In the non-Hermitian AA limit, the system undergoes a delocalization transition by tuning the potential strength. At the critical point, the spatial distribution of the critical state shows a self-similar structure with the relative distance between the peaks being the Fibonacci sequence,
and the finite-size scaling of the inverse participation ratios $({\rm IPRs})$ of the critical ground state with lattice size $L$ shows that ${\rm IPR}_g\propto L^{-0.1189}$.
In the non-Hermitian Fibonacci limit, we find that the system is always in the extended phase.
Along the continuous deformation from the non-Hermitian AA model into the non-Hermitian Fibonacci model in the IAAF model,
the cascade of the delocalization transition is found, but only a few plateaux appear.
Moreover, the self-similar structure of spatial distribution for the critical modes along the cascade transition is also found.
In addition, we find that the delocalization transition and the real-complex transition for the excited states happen at almost the same parameter.
Our results show that the non-Hermiticity provides an additional knob to control the cascade of the delocalization transition besides the on-site potential.
\end{abstract}

\maketitle
\section{\label{intro}Introduction}
The quasiperiodicity is the abstraction of matter with short-range order but without long-range order.
In a quasiperiodic matter, unlike an ordered system, the translational invariance is broken by the incommensurate period, but unlike a disordered system, the long-range correlation still persists.
These special features make the quasiperiodic system not only inherit the physics of both ordered and disordered systems, such as Anderson localization
~\cite{Aubry1980,Schiffer2021,Lahini2009,Hiramo1989},
but also exhibit lots of novel phenomena, such as the fractal eigenmodes~\cite{Jitomirskaya1999,Yao2019,Mace2017,Agrawal2020}.

Many theoretical quasi-periodic models~\cite{Aubry1980,Schiffer2021,Lahini2009,Hiramo1989,Jitomirskaya1999,Mace2017},
 including the bichromatic lattices~\cite{Yao2019},
electronic materials in orthogonal magnetic field~\cite{Hofstadter1976}, have been proposed to study the delocalization transition
and the critical behavior.
In particular, many one-dimensional models have been intensively studied due to its simplicity and experimental realization~\cite{Ganeshan2013,Mastropietro2015,Xu2019,Sinha2019,Xu2021,Kraus2012_1,Zeng2020,Zeng2020_2,Verbin2013,Merlin1985,Macia1999,Ashraff1989,Goblot2020}.
Among these models, the Aubry-Andr\'e model (AA)~\cite{Aubry1980,Schiffer2021,Ganeshan2013,Mastropietro2015,Xu2019,Sinha2019,Xu2021,Zeng2020,Zeng2020_2}
and the Fibonacci model~\cite{Mace2017,Merlin1985,Macia1999,Ashraff1989} are two of the most celebrated examples.
For the AA model, the quasiperiodicity enters in the form of an on-site cosine modulation incommensurate with the underlying periodic lattice spacing,
and the delocalization transition occurs at a critical value of the quasiperiodic potential~\cite{Aubry1980}.
For the Fibonacci model, the potential is a binary chain, and it has a modulation with two discrete values that appear interchangeably according to the Fibonacci sequence~\cite{Mace2017,Merlin1985,Macia1999,Ashraff1989}.
Theoretical and experimental studies have shown that the Fibonacci model always has critical wavefunctions for any values of quasi-periodic potential~\cite{Mace2017,Merlin1985,Macia1999,Ashraff1989,Goblot2020}.
Moreover, recently it was shown that many exotic properties appear in the interpolating Aubry-Andr{\'e}-Fibonacci (IAAF) model, which combines the AA model and the Fibonacci model~\cite{Goblot2020,Kraus2012,Verbin2013}.
Based on the IAAF model, the AA model and the Fibonacci model share the same topological properties and belong to the same topological class~\cite{Kraus2012}.
Despite these two limits, there is a wide range of parameter space unexplored in the IAAF model.
Together with theory and experiment, Ref.~\cite{Goblot2020} finds that a cascade of delocalization transition occurs when the interpolating
parameter runs from the limit of AA model to the limit of the Fibonacci model.

On the other hand, the delocalization transition is also found in the non-Hermitian disordered or quasi-periodic systems
~\cite{Hatano1996,Hatano1997,Hatano1998,Kolesnikov2000,Longhi2019,Longhi20192,Cai2021,Jiang2019,Liu2021,Jazaeri2001,Feinberg1999,PWang2019,Hamizaki2019,zhai2020,Liu2020_1,Tang2021,Cai20212}.
Due to the releasing of Hermiticity constrain, the non-Hermitian systems exhibit much richer phenomena than their Hermitian counterparts~\cite{Yang2020,Xu2017,Kunst2018,Gong2018,Yao20180,zhai20200}, such as the topological non-Hermitian skin effect under open boundary condition (OBC)~\cite{Yao2018,Song2019,Okuma2020,Kawabata2018,Borgnia2020,Longhi2020,Fu2021,Liu2020,Alvarez2018,Zhang2020,Yang20203,Yi2020,Yoshida2020},                                                                                                                                            exceptional points~\cite{Yoshida2018,Kawabata2019,Yin2017,Dora2019,Ding2016}, etc.
Many interesting critical behaviors were found in the non-Hermitian systems~\cite{Li2020,zhai2018,Bender1998,Bender2007},
and the fundamental concepts in the usual critical systems, such as, the band gaps and locality, have been challenged~\cite{Bender1998,Bender2007}.
Moreover, effects induced by the non-Hermiticity in the delocalization transition have been studied in different contexts~\cite{Longhi2019,Longhi20192,Cai2021,Jiang2019,Liu2021,Jazaeri2001,Feinberg1999,PWang2019,Hamizaki2019,zhai2020}.

Here, we investigate the effect of non-Hermiticity in the cascade of the delocalization transition in the IAAF model. By introducing the non-reciprocal hopping term, we construct a non-Hermitian IAAF model.
In the non-Hermitian AA limit, the system shows a delocalization transition by tuning the strength of the quasiperiodic potential,
while in the non-Hermitian Fibonacci model limit, we find this model is always in the extended phase.
Along the continuous deformation from the non-Hermitian AA limit into the non-Hermitian Fibonacci limit,
the cascade of the inverse participation ratios (${\rm IPRs}$) is found, similar to its Hermitian counterpart. However, we find that for the non-Hermitian IAAF model, there are only a few plateaux. This is quite different from the Hermitian case. Moreover, the critical properties of the delocalization are also studied, and the correspondence between the delocalization transition and the real-complex transition is verified.
Our results demonstrate that the non-Hermiticity provides an additional knob to control the cascade of the delocalization transition.

The remainder of the paper is organized as follows. In Sec.~\ref{secmodel}, the non-Hermitian IAAF model is presented.
In Sec.~\ref{secResultI}, the delocalization transition and critical behavior of the non-Hermitian AA limit and the non-Hermitian Fibonacci limit are studied.
Then the cascade of delocalization transition along the continuous deformation from the non-Hermitian AA model into the non-Hermitian Fibonacci model is explored in Sec.~\ref{secResultII}.
A summary is given in Sec.~\ref{secSum}.

\section{\label{secmodel}The non-Hermitian IAAF Model}
In this paper, the non-Hermiticity is induced by the non-reciprocal hopping. The non-Hermitian IAAF model then reads
\begin{eqnarray}
\label{Eq:model}
H=\sum_{j=1}^{L}(te^{g}c^{+}_{j+1}c_{j}+te^{-g}c^{+}_{j}c_{j+1})+\lambda V_{j}(\beta)c^{+}_{j}c_{j},
  \end{eqnarray}
where $c_j^{+}(c_j)$ are creation (annihilation) operators at site $j$, $\lambda$ measures the strength of the on-site potential, and $L$ is the lattice size.
$t$ and $g$ label the non-reciprocal hopping between nearest-neighbour sites. In the following, we assume $t=1$ as the unit of energy.

The on-site potential $V_j(\beta)$ is written as~\cite{Goblot2020}
\begin{eqnarray}
  V_j(\beta) &=& -\frac{\tanh[\beta\left(\cos(2\pi \alpha j+\phi)-\cos(\pi\alpha)\right)]}{\tanh{\beta}},
\end{eqnarray}
where $\phi$ is a random phase, and $\beta$ is a tunable parameter,
and $\alpha$ is an irrational spatial modulation frequency.
For the infinity system $\alpha$ is usually chosen to be the inverse golden ratio $(\sqrt5-1)/2$, which can be approached by $\alpha=\lim_{n\rightarrow\infty}{F_n/F_{n+1}}$
with $F_n$ being the \emph{n}th Fibonacci number.
For the finite system, the potential must be periodic for the periodic boundary condition (PBC),
hence, $\alpha$ has to be approximated by a rational number $F_{n}/F_{n+1}$
with site number $L=F_{n+1}$.

\begin{figure}
  \centering
  \includegraphics[width=3 in]{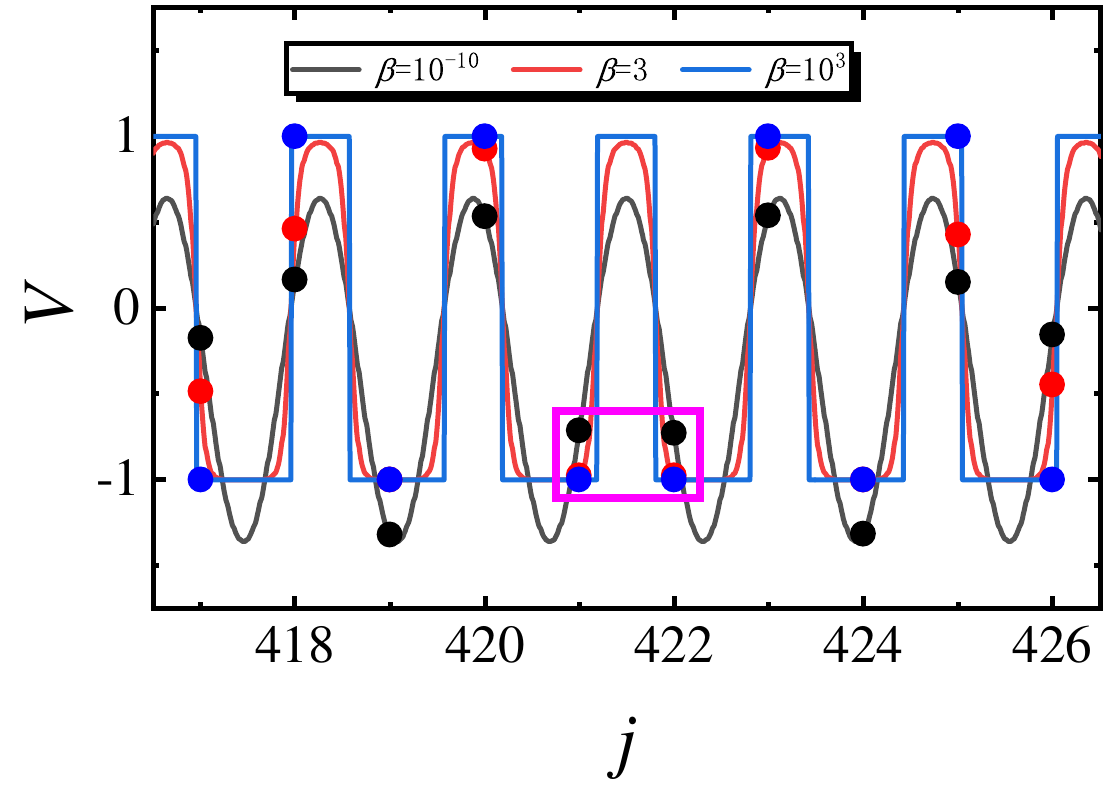}\\
  \caption{Evolution of the spatial on-site potential for several values of $\beta$.
  The discrete values of $V_j(\beta)$ are sampled.
  Here, we set $\phi=0$.}
  \label{fig:Vj}
\end{figure}

For very small $\beta$, like $\beta=10^{-10}$, the on-site potential $V_j(\beta)$ reduces to $V_j=\cos{(2\pi\alpha j+\phi)}-\cos{(\pi b)}$,
which is the AA modulation with a constant energy shift.
With the increase of $\beta$, the continuous function of $V_{j}(\beta)$ becomes steeper as shown in Fig.~\ref{fig:Vj},
and the range of possible values of $V_{j}$ shrinks.
For very large $\beta$, like $\beta=10^{3}$, $V_{j}(\beta)$ becomes a step potential switching between $\pm 1$ according to the Fibonacci sequence.
As a result, the model~(\ref{Eq:model}) can be continuously changed from the AA model to the Fibonacci model by tuning $\beta$.

\section{\label{secResultI}The delocalization transition and critical behavior in the non-Hermitian AA and the non-Hermitian Fibonacci limit}
\subsection{The non-Hermitian AA limit}
\begin{figure*}
  \centering
  \includegraphics[width=4 in]{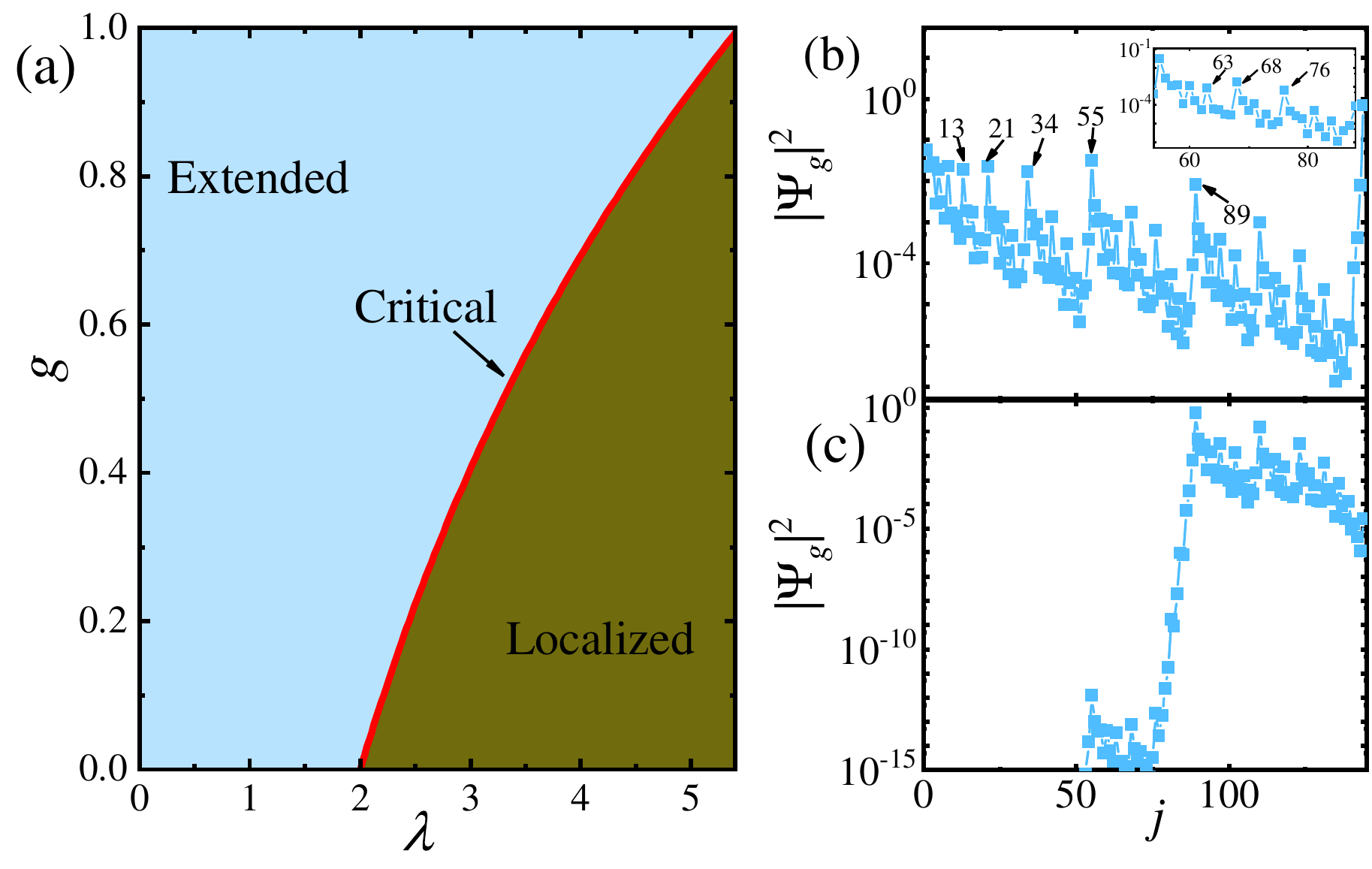}\\
  \caption{(a) Phase diagram of the non-Hermitian AA model. The blue region is the extended phase, and the yellow region is localized phase,
   and the red line separating these two regions is the critical line.
   Typical spatial distributions of the critical state under PBC and OBC are shown in (b) and (c), respectively.
   In (b) and (c), we set $\phi=0$, $L=144$ and $\beta=10^{-10}$ in the numerical calculation.
   Some location of the peak values are labeled, and $\Psi_g$ is the ground state according to the real parts of the eigenenergies.
   The insert in (b) is the spatial distribution between $j=55$ and $88$ with the some condition with (b).}
  \label{fig:AAPHASE}
\end{figure*}

For the Hermitian AA model, all states are extended for $\lambda/t<2$, and they are localized for $\lambda/t>2$,
and at $\lambda/t=2$ all states are critical~\cite{Aubry1980}.
It was shown that the non-reciprocal hopping changes the critical point between the extended phases and the localization phase to be~\cite{Jiang2019}
\begin{eqnarray}
  \lambda_c &=& 2te^{g}.
\end{eqnarray}
The phase diagram ($g>0$) of the non-Hermitian AA model is given in Fig.~\ref{fig:AAPHASE} (a).
When $\lambda<2te^{g}$, the non-Hermitian AA model is in the extended phase.
In this phase, the system has an edge state under the OBC due to the non-Hermitian skin effect~\cite{Yao2018}.
When $\lambda>2te^{g}$, the non-Hermitian AA model is in the localized phase,
and the localized state has an asymmetrical exponential decay under both the PBC and OBC,
which demonstrates different localization lengths on different sides from the localization center.
Moreover, it has been demonstrated that $\lambda_c=2te^{g}$ is also the boundary between the topologically trivial and non-trivial phases~\cite{Jiang2019}, i.e.,
the localization phase is also the topological trivial phase with zero winding number,
and the extended phase corresponds to the topological non-trivial phase with winding number being $1$ for $g>0$.
Due to the bulk-bulk correspondence~\cite{Jiang2019}, the extended phase of $g>0$ should have a right-skin edge state under OBC.

%On the boundary of $\lambda_c=2te^{g}$, all states do not localized on any sites, nor have a distinct edge state under the OBC, which are the critical states.
At $\lambda_c=2te^{g}$, the states of the system are all critical.
A significant feature of the critical state is the self-similar behavior in the spatial distribution.
Here, we use the right ground state $\Psi_g$ in the critical phase with $L=144$ and $\phi=0$ as an example to illustrate the self-similar structure.
It should be noted that the ordering of the eigenstates is according to the real parts of the eigenenergies,
that is, the ground state corresponds to the eigenstate with lowest real part of eigenenergy.
As shown in the Fig.~\ref{fig:AAPHASE} (b), the typical spatial distribution of the critical ground state under PBC is plotted.
It is found that the peaks of the spatial distribution always satisfy the Fibonacci sequence, i.e.,
the locations of peak values are $1, 2, 3 \cdots, 55, 89, 144$ as labeled in Fig.~\ref{fig:AAPHASE} (b).
Moreover, the relative distance between secondary peaks located between the primary peaks also satisfy the Fibonacci sequence,
for example, the relative distance between the secondary peaks ranging from $55$ to $89$ also satisfy the Fibonacci sequence as shown in the insert of Fig.~\ref{fig:AAPHASE} (b).
This self-similar structure is also verified for other lattice size and different $\phi$.
In Fig.~\ref{fig:AAPHASE} (c), the spatial distribution of the critical state under the OBC is plotted.
One finds that the wave function is localized near the right side. But different from the skin effect, the wave function is not localized on the boundary.
\begin{figure}
  \centering
  \includegraphics[width=3.5 in]{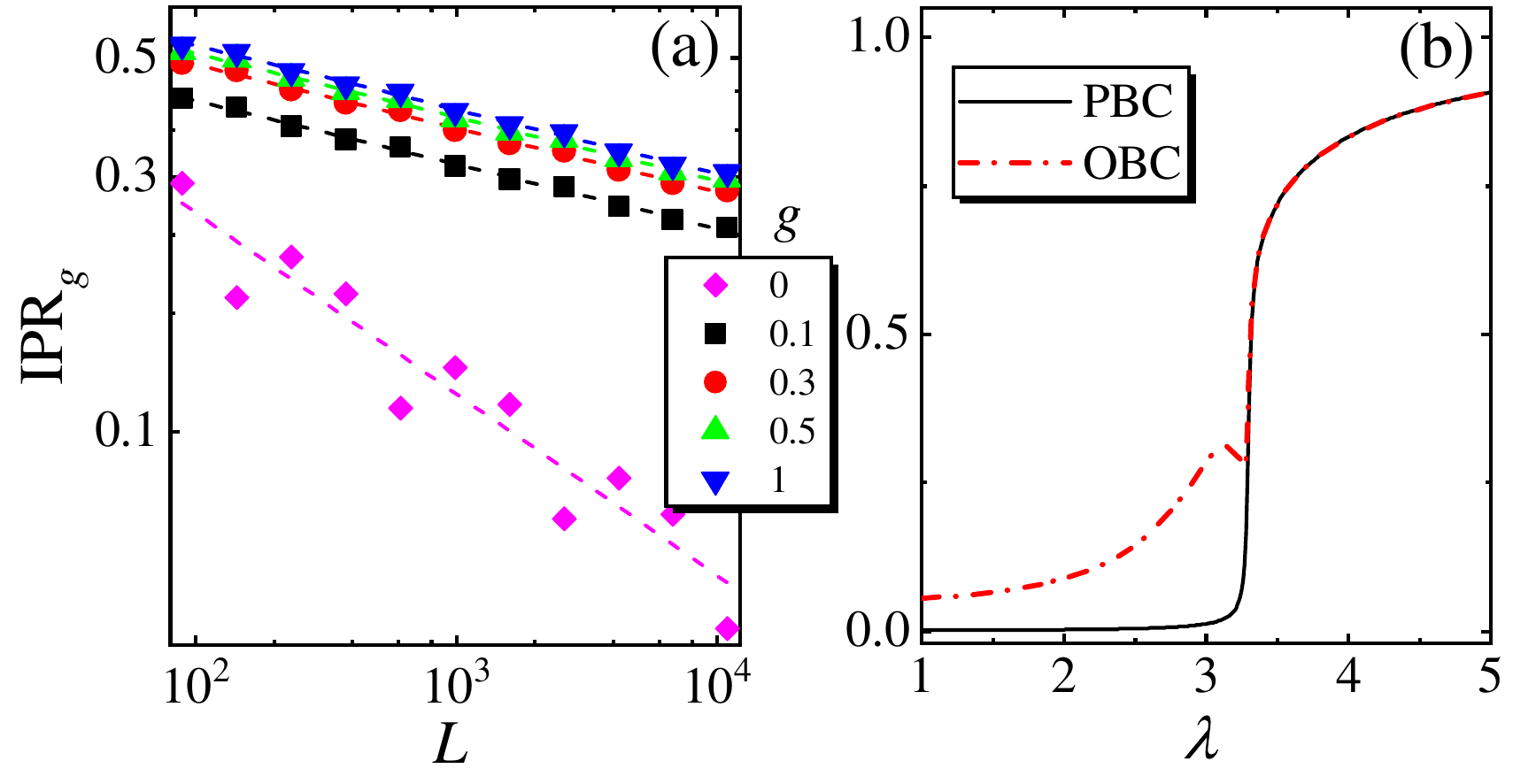}\\
  \caption{(a) Finite-size scaling of ${\rm IPR}_g$ of the critical ground state
  with lattice size $L$ under PBC for different $g$ and the fitted lines (dashed lines) for $\beta=10^{-10}$ (that is, in the non-Hermitian AA limit).
  (b) ${\rm IPR}_g$ v.s. $\lambda$ for $g=0.5$ under PBC and OBC  for $\beta=10^{-10}$ and $L=610$.}
  \label{fig:AAIPR}
\end{figure}
\begin{figure*}
  \centering
  \includegraphics[width=5 in]{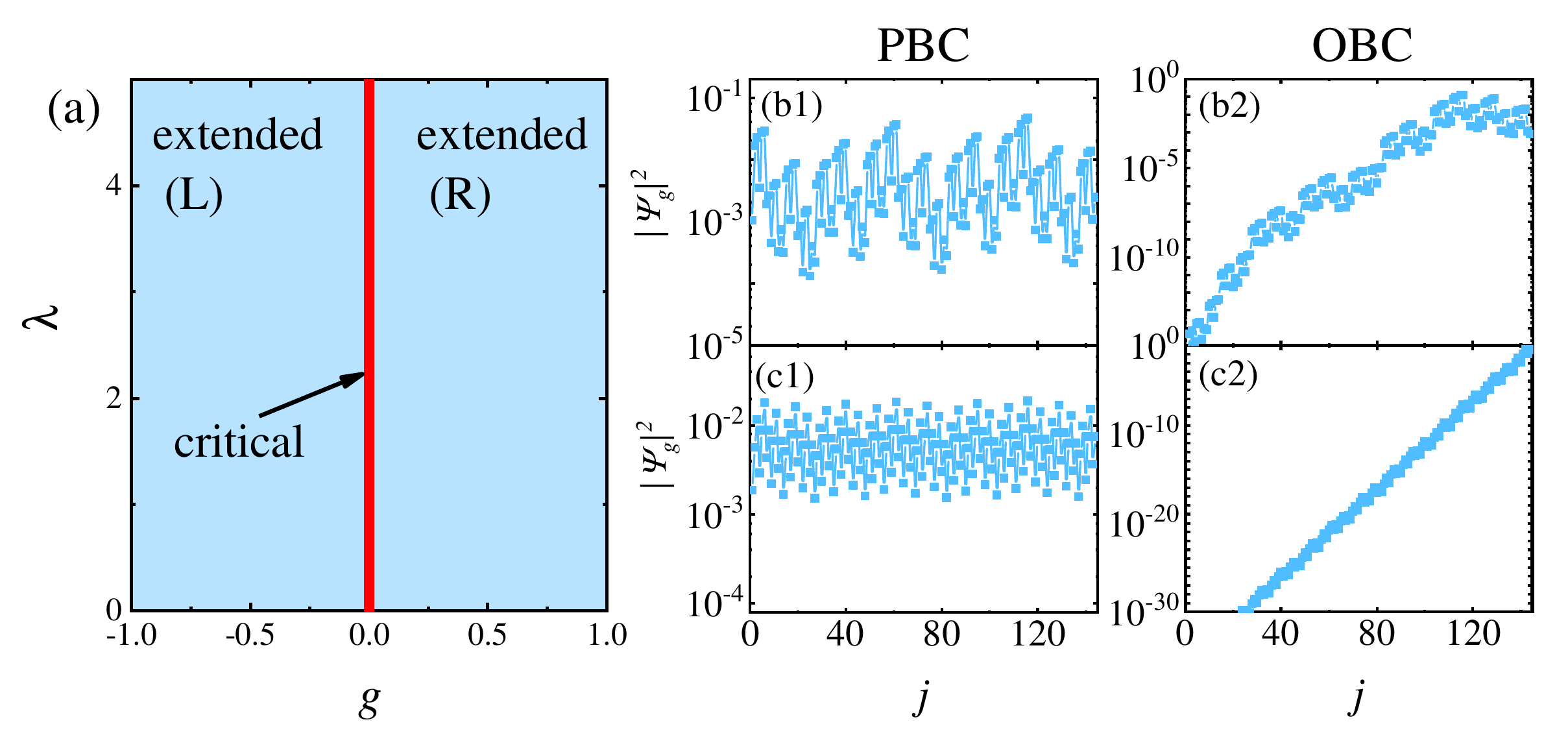}\\
  \caption{(a) The phase diagram of the non-Hermitian Fibonacci model. \{L, R\} represent the left-skin and right-skin extended phase.
  Typical spatial distributions for eigenmodes of the non-Hermitian Fibonacci model PBC and OBC are shown in (b1), (b2), (c1) and (c2).
  Here, we use $L=144$, $\lambda=5$, $\beta=10^{10}$ and $g=0.1$ for (b1) and (b2) and $g=2$ for (c1) and (c2).}
  \label{fig:FibPhase}
\end{figure*}

To further study the behavior of the critical mode,
we calculate the $\rm IPR$ of the right eigenstate $\Psi_{n}$ of the Hamiltonian~\cite{Goblot2020,Liu2021},
\begin{eqnarray}
  {\rm IPR}_n &=& \frac{\sum_{j=1}^{L}{|\Psi_{n,j}|^4}}{\sum_{j=1}^{L}{|\Psi_{n,j}|^2}},
\end{eqnarray}
where $n$ labels the $n$th eigenstate of system according to the real part of the eigenvalues.
${\rm IPR}_n$ is usually used to detect the delocalization transition in both the Hermitian and non-Hermitian systems~\cite{Evers2000,Cuevas2002,Jiang2019,Liu2021}.
For the extended state, the wave function is homogeneously distributed through all sites,
and ${\rm IPR}_n$ scales with $L$ as ${\rm IPR}_n\propto L^{-1}$.
On the other hand, ${\rm IPR}_n$ scales as ${\rm IPR}_n\propto L^{0}$ for the localized mode~\cite{Goblot2020,Evers2000}.

Moreover, it was shown that the IPR of the critical mode also satisfies a power law with respect to $L$~\cite{Evers2000,Cuevas2002}.
As shown in Fig.~\ref{fig:AAIPR} (a), the $L$ dependence of IPR of the critical ground state
$({\rm{IPR}}_g)$ for $\phi=0$ and different $g$ are plotted.
It is found that curves of ${\rm{IPR}}_g$ versus $L$ are parallel straight lines in the log-log scale,
which demonstrates that $\rm{IPR}_g$ scales as ${\rm{IPR}}_g\propto L^{\nu}$ for any $g\neq0$.
By a linear fitting, the average $\nu$ is found to be $\nu=-0.1189$.
Moreover, the ${\rm{IPR}}_g$ v.s. $L$ and the fitted line for $g=0$ is also plotted in Fig.~\ref{fig:AAIPR} (a) as a comparison,
and the fitted result shows $\nu=-0.2539$, which demonstrates that the non-Hermitian and Hermitian AA models belong to different universal classes.

The averaged $\rm IPR_g$ as a function of $\lambda$ under the PBC and OBC are plotted in Fig.~\ref{fig:AAIPR} (b).
It is shown that in the regime of extended phase, the ${\rm IPR}_g$ under OBC is larger than that under PBC,
due to the boundary-localization nature of OBC~\cite{Yao2018,Jiang2019}.
In the regime of localized phase, the ${\rm IPR}_g$ under both the PBC and OBC are almost the same.
In addition, a minimum of ${\rm IPR}_g$ under OBC is found at $\lambda\approx 3.27$,
which is close to the theoretical predict value of $2e^{0.5}\approx3.29$. The small discrepancy comes from the finite-size effects.
Hence, the minimum of ${\rm IPR}_g$ under OBC can be used to determine the location of the critical state numerically.

\subsection{The non-Hermitian Fibonacci limit}

At $g=0$, it is well known that all the eigenstates are always critical at any $\lambda/t>0$,
and the spatial distribution of the critical mode has a self-similar structure~\cite{Macia1999,Goblot2020}.
However, for the non-Hermitian Fibonacci model, we find that the eigenstates are in the extended phase,
and the self-similar structure is destroyed by the non-Hermiticity.

The phase diagram of the non-Hermitian Fibonacci model is sketched in Fig.~\ref{fig:FibPhase} (a). The typical spatial distributions of the eigenstates with different $g$
under PBC and OBC are plotted in Figs.~\ref{fig:FibPhase} (b1), (b2), (c1) and (c2). Under the PBC, the spatial distribution of the ground state
has a self-similar structure for small $g$, but it tilts owing to the effect of the non-reciprocal hopping, as shown in Fig.~\ref{fig:FibPhase} (b1) for $g=0.1$.
In contrast, for larger $g$, the self-similar structure fades away, as plotted in Fig.~\ref{fig:FibPhase} (c1) for $g=2$.
In addition, the difference of the distribution on different sites becomes smaller.
Under OBC, as a result of the skin effect~\cite{Jiang2019}, the edge states appear on the right boundary for $g>0$ as shown in Figs.~\ref{fig:FibPhase} (b2) and (c2).
For $g<0$, similar behaviors of spatial distribution can also be found, but the edge state should be localized on the left boundary.
Apparently, the non-Hermitian Fibonacci model should undergo a transition between the left-skin extended phase and right-skin extended phase,
when $g$ varies from negative to positive.

\begin{figure}
  \centering
  \includegraphics[width=3 in]{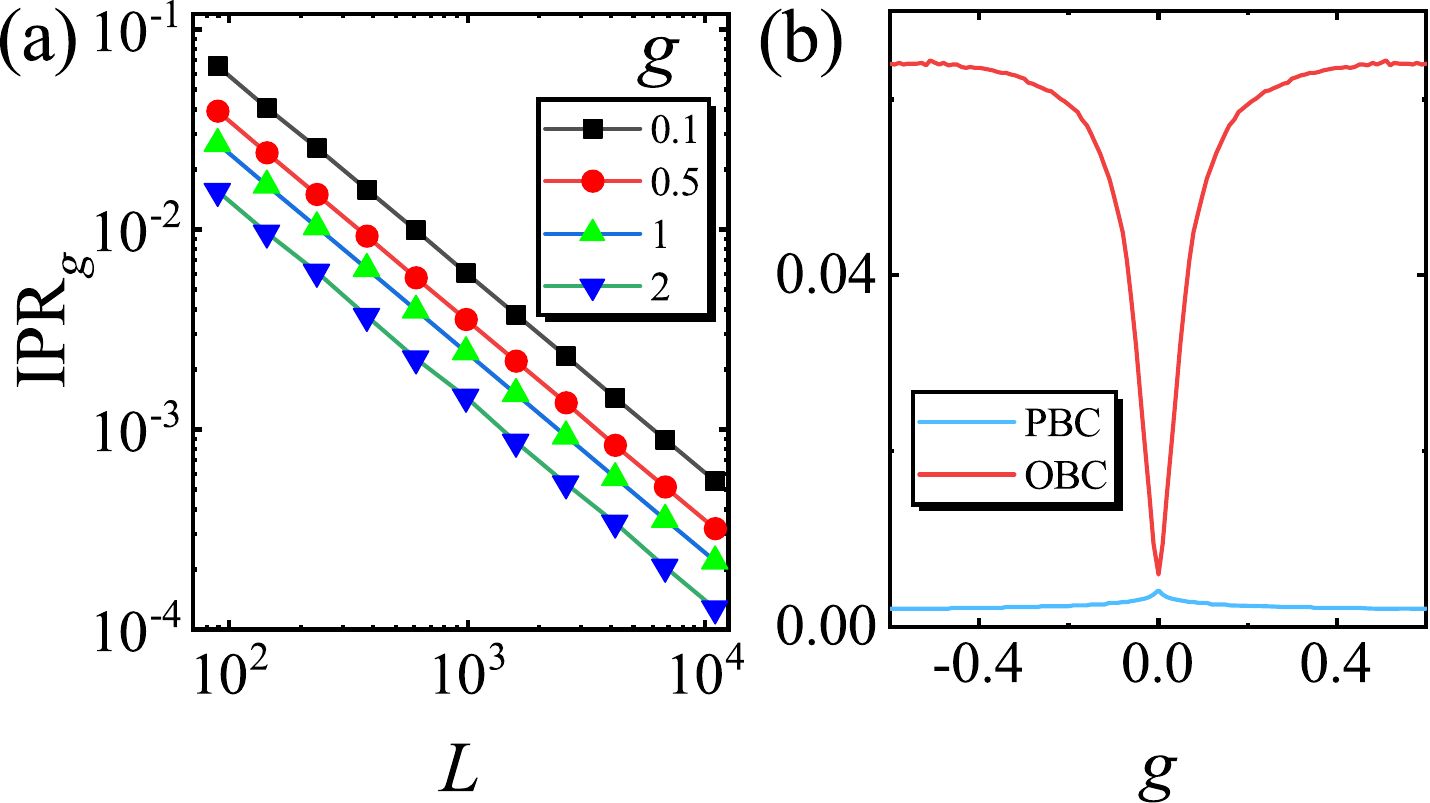}\\
  \caption{(a) ${\rm{IPR}}_g$ versus $L$ for different $g$ and $\beta=10^{10}$ ( that is, in the non-Hermitian Fibonacci limit).
  (b) ${\rm IPR}_g$ versus $g$ under PBC and OBC for $\beta=10^{10}$.
  $\lambda=5$ is used in the numerical calculation, and the lattice size is $L=610$ in (b).
  Both of the results in (a) and (b) are averaged for 500 choices of $\phi$.}
  \label{fig:FibIPRg}
\end{figure}

In Fig.~\ref{fig:FibIPRg}(a), the behavior of ${\rm{IPR}}_g$ versus $L$ is studied for different $g$.
It is found that ${\rm{IPR}}_g$ scales as ${\rm{IPR}}_g\propto L^{\nu}$ for different $g$.
A power-law fitting shows that the averaged $\nu$ is $-0.9932$ close to $-1$, which means that
the non-Hermitian Fibonacci model is in the extended phase for any $g\neq0$.
The ${\rm{IPR}}_g$ under PBC and OBC as a function of $g$ are also plotted in Fig.~\ref{fig:FibIPRg} (b),
and the minimum of ${\rm{IPR}}_g$ under OBC and the peak of ${\rm{IPR}}_g$ under PBC are found at $g=0$,
indicating that the Hermitian Fibonacci model is in the critical phase separating the left-skin and right-skin phases.

\section{\label{secResultII}Cascade of IPRs along the transition from AA model into Fibonacci model}
Here, we study the delocalization in the non-Hermitian IAAF model~(\ref{Eq:model}).

\subsection{Ground state}
\begin{figure}
  \centering
  \includegraphics[width=3.5 in]{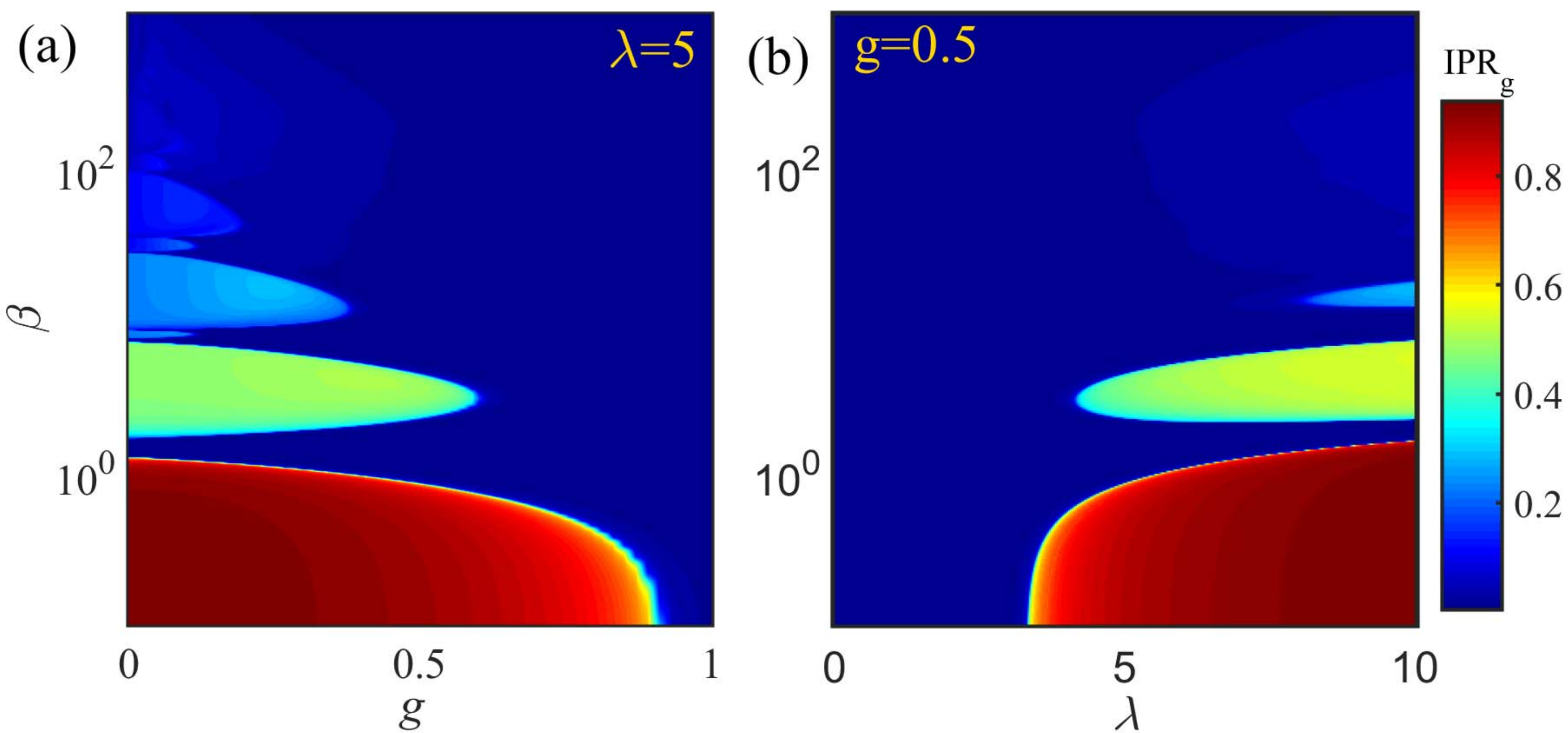}\\
  \caption{(a) ${\rm IPR}_g$ under PBC as a function of $\beta$ and $g$ for $\lambda=5$, and (b) ${\rm IPR}_g$ under PBC as a function of $\beta$ and $\lambda$ for $g=0.5$.
  The numerical result is determined for $L=610$, and ${\rm{IPR}}_g$ is averaged for 100 choices of $\phi$.}
  \label{fig:IAAFPHASE}
\end{figure}

First, we show the cascade of $\rm IPR$ in the ground state.
In Fig. \ref{fig:IAAFPHASE} (a), the ${\rm IPR}_g$ versus $\beta$ and $g$ for $\lambda=5$ is plotted.
For small $g$, ${\rm IPR}_g$ shows a cascade behavior, in which the lobes of localization regions with large ${\rm IPR}_g$ are separated by the delocalization transitions with the minima of ${\rm IPR}_g$.
In addition, for large $\beta$, the localized regions shrink. This is similar to the the Hermitian case~\cite{Goblot2020}. However, for large $g$, one finds that the number of plateaux of ${\rm IPR}_g$ becomes small.
Such behaviors of ${\rm IPR}_g$ means that the non-reciprocal hopping tends to destroy the localization.
On the other hand, the ${\rm IPR}_g$ under PBC as a function of $\beta$ and $\lambda$ at $g=0.5$ is plotted in Fig.~\ref{fig:IAAFPHASE} (b).
The cascade behavior also manifests itself in this case. For larger $\lambda$, the number of plateaux increases. From Figs.~\ref{fig:IAAFPHASE} (a) and (b), one finds that the cascade behavior in the non-Hermitian IAAF
model can be tuned by $g$. For small $g$ and large $\lambda$, the behavior is similar to the the Hermitian case~\cite{Goblot2020} when $\beta$ is small. But for large $\beta$, the cascade behavior disappear. This is quite different from the Hermitian case, in which the cascade will continue to $\beta\rightarrow\infty$ as long as the resolution for the ${\rm IPR}_g$ is high enough.

The mechanism of the cascade behavior is similar to the Hermitian case~\cite{Goblot2020}.
In the AA model limit, the ground eigenstates of the strong localized mode are always localized at a single site, and the value of ${\rm{IPR}}_g$ is almost $1$.
With the increase of $\beta$, the potential of the two-site pair, two neighbor sites having almost the same potential, goes down towards the minimum of the potential,
e.g., the sites $421$ and $422$ labeled in Fig.~\ref{fig:Vj}.
Thus, the energy of the two-site localized state decreases and the two-site localized state becomes the new ground state~\cite{Goblot2020}.
In the transition region between the single site localization and two-site localization, the system is in the extended phase.
The ${\rm{IPR}}_g$ decreases suddenly once the system enters into the transition region from the localized phase,
and then increases when the two-site localized states become the ground states.
%Similarly, with the further increase of $\beta$, the potential values of four-site pairs and eight-site pairs will
%become the lowest of the potential successively, and the localized states on these pairs turn into the ground sates.
Similarly, with the further increase of beta, the potential values of higher-rate pairs, like the four-, eight-, and so on, will become the lowest potential,
and the corresponding localized states turn into the ground state successively.
Therefore, the similar structure of ${\rm{IPR}}_g$ appears again.
As a result, the ${\rm{IPR}}_g$ shows the cascade behavior with the increase of $\beta$. However, different from the Hermitian case, there is only a few plateaux in the non-Hermitian IAAF model. To explore the reason,
we find that for the Hermitian IAAF model, by increasing $\beta\rightarrow\infty$, the $\rm IPR$ will display a series of plateaux
corresponding to the two-sites localization mode, four-sites localization mode, eight-sites localization mode, and so on.
Therefore, the localized modes gradually extend to critical in the Fibonacci limit, where the eigenstates are self-similar~\cite{Goblot2020}.
But the non-Hermitian Fibonacci model is in the extended states. The self-similarity is truncated for some energy levels. This makes the number of the plateaux limited to a small value.

To further explore the cascade of ${\rm IPR}_g$, the ${\rm{IPR}}_g$ versus $\beta$ under PBC and OBC for some fixed values of $\lambda$ and $g$ are shown in Fig.~\ref{fig:betaIPR}.
For small $\beta$, the values of the ${\rm{IPR}}_g$ with the PBC and the OBC coincide in the localized plateaux. However, in the delocalization transition regions, the ${\rm{IPR}}_g$ under OBC is larger than that under PBC.
The reason is that for the OBC, the ${\rm IPR}_g$ always have a larger finite value as shown in Figs.~{\ref{fig:AAIPR} and \ref{fig:FibIPRg}, as a result of the skin effect. In addition, Fig.~\ref{fig:betaIPR} (a) shows that, for $\lambda=5$, the value of the ${\rm IPR}_g$ in the first plateaux decreases as $g$ increases. The reason is that for large $g$ the hopping is enhanced while the localization is suppressed. For the same reason, the transition region between two plateaux is much broader for larger $g$. With the increase of $\lambda$, one finds that first plateaux of $g=0.5$
and $g=0.1$ overlaps with each other and the transition region shrinks, as shown in Figs.~\ref{fig:betaIPR} (b), (c) and (d).

\begin{figure}
  \centering
  \includegraphics[width=3 in]{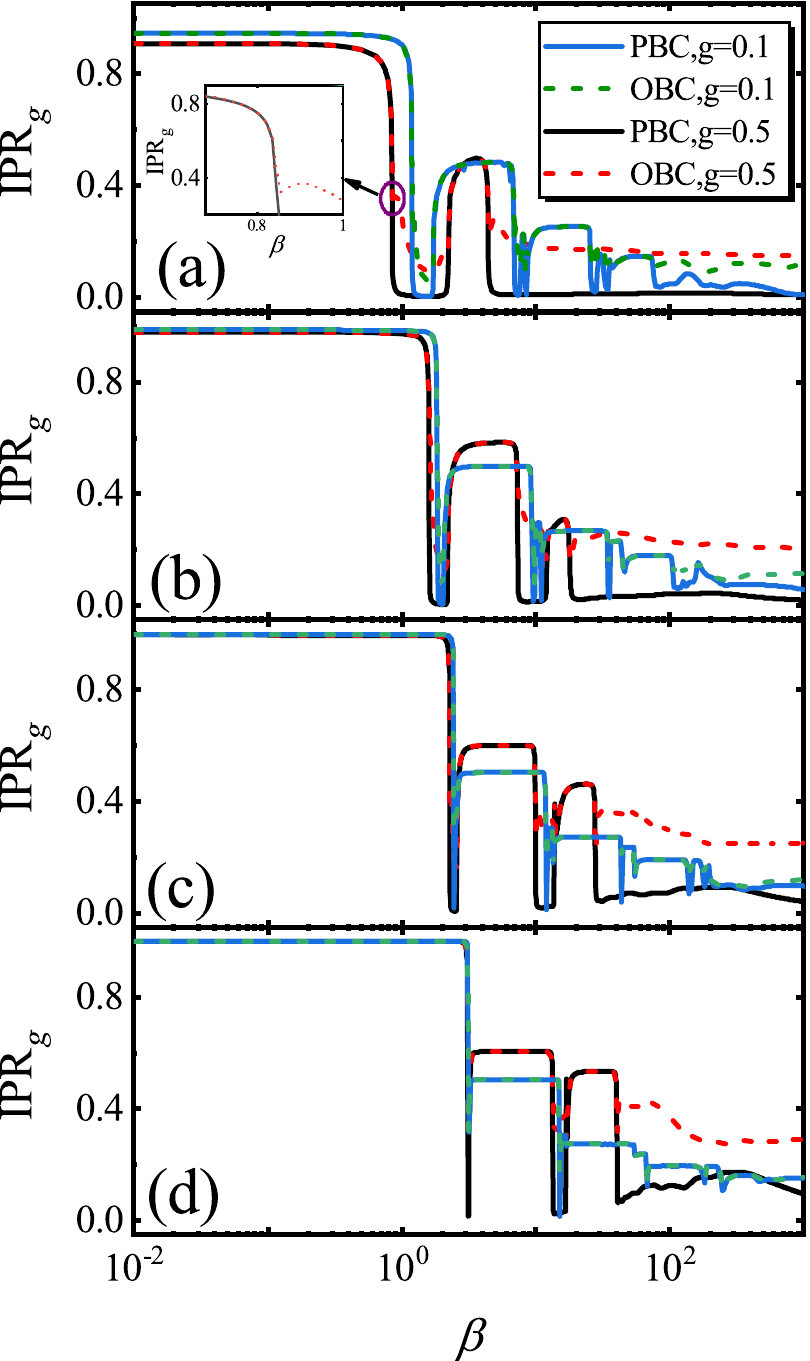}\\
  \caption{${\rm{IPR}}_g$ as a function of $\beta$ for different $\lambda$ and $g$ under PBC and OBC.
  Here, we use $\lambda=5$ in (a), $\lambda=10$ in (b), $\lambda=20$ in (c) and $\lambda=50$ in (d), and $L=610$.
  The result is averaged for 500 choices of $\phi$.
  The inset in (a) shows the location of critical mode between the first plateaux and the transition region.}
  \label{fig:betaIPR}
\end{figure}

Another interesting feature of the cascade of ${\rm{IPR}}_g$ is that the ${\rm IPR}_g$ in the subsequent plateaux (if exist) becomes larger for larger $g$ for any $\lambda$, in contrast to the case for the first plateau.
As noted above, these plateaux corresponds to the ground state with the two or higher-site localized modes. For the Hermitian case, the eigenmode is equably distributed in these sites~\cite{Goblot2020}.
However, for the non-Hermitian Hamiltonian, the distribution weight is different for different sites~\cite{Jiang2019}.
As shown in Fig.~\ref{fig:twoletter},
the spatial distribution of ground states of the two-site localized modes for $g=0, 0.1$ and 0.5 are plotted.
Here, we use $\lambda=10$, $\phi=0$ and $\beta=5$, and the sites $421$ and $422$ are the two-site pair with lowest potential.
It is clear the eigenstate is equally distributed on these two sites for $g=0$. However, with the growth of $g$, the distribution weight on the right site $422$ becomes greater than that of the left site $421$ owing to the non-reciprocal hopping.
That is, the distribution in one site will dominate the two-site localized state in the non-Hermitian case.
As a result, the ${\rm{IPR}}_g$ will increase with $g$ for the higher-site localized states.
\begin{figure}
  \centering
  \includegraphics[width=3 in]{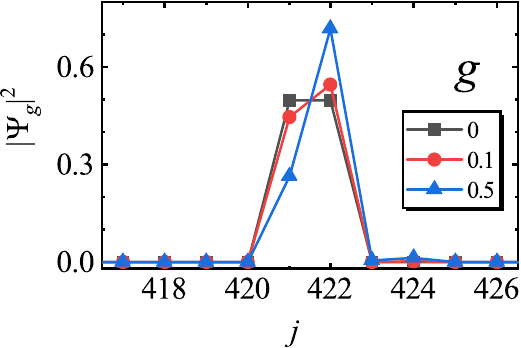}\\
  \caption{Typical spatial distribution of the two-site localization state with $g=0$, $g=0.1$ and $g=0.5$.
  Here, we use $\lambda=10$, $\beta=5$, $L=610$ and $\phi=0$.}
  \label{fig:twoletter}
\end{figure}

\begin{figure}
  \centering
  \includegraphics[width=3 in]{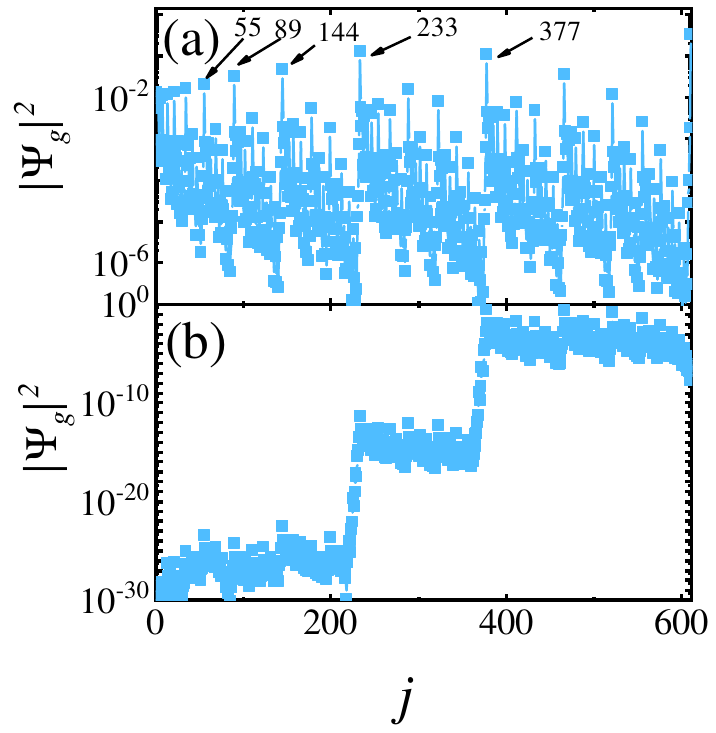}\\
  \caption{Typical spatial distribution of critical mode along the transition from AA model to Fibonacci model
  under PBC (a) and OBC (b).
  Here, we use $\lambda=5$, $\beta=0.85$, $L=610$, $g=0.5$ and $\phi=0$, which locates between the first and
  the transition region in Fig.~\ref{fig:betaIPR} (a) (corresponding to the minimum of ${\rm IPR}_g$ under OBC
  shown in the insert in Fig.~\ref{fig:betaIPR} (a)).}
  \label{fig:beta085}
\end{figure}
Moreover, one finds that ${\rm{IPR}}_g$ under OBC has a minimum between the plateaux and the transition region, as shown in the insert of Fig.~\ref{fig:betaIPR} (a).
This indicates the appearance of the critical mode.
For these critical modes, we find that the self-similar structure is still preserved under the PBC as shown in Fig.~\ref{fig:beta085} (a). In addition, for the OBC, the wave function is distributed in one side but not at the boundary, as shown in Fig.~\ref{fig:beta085} (b).

\subsection{Excited states}

\begin{figure}
  \centering
  \includegraphics[width=3.5 in]{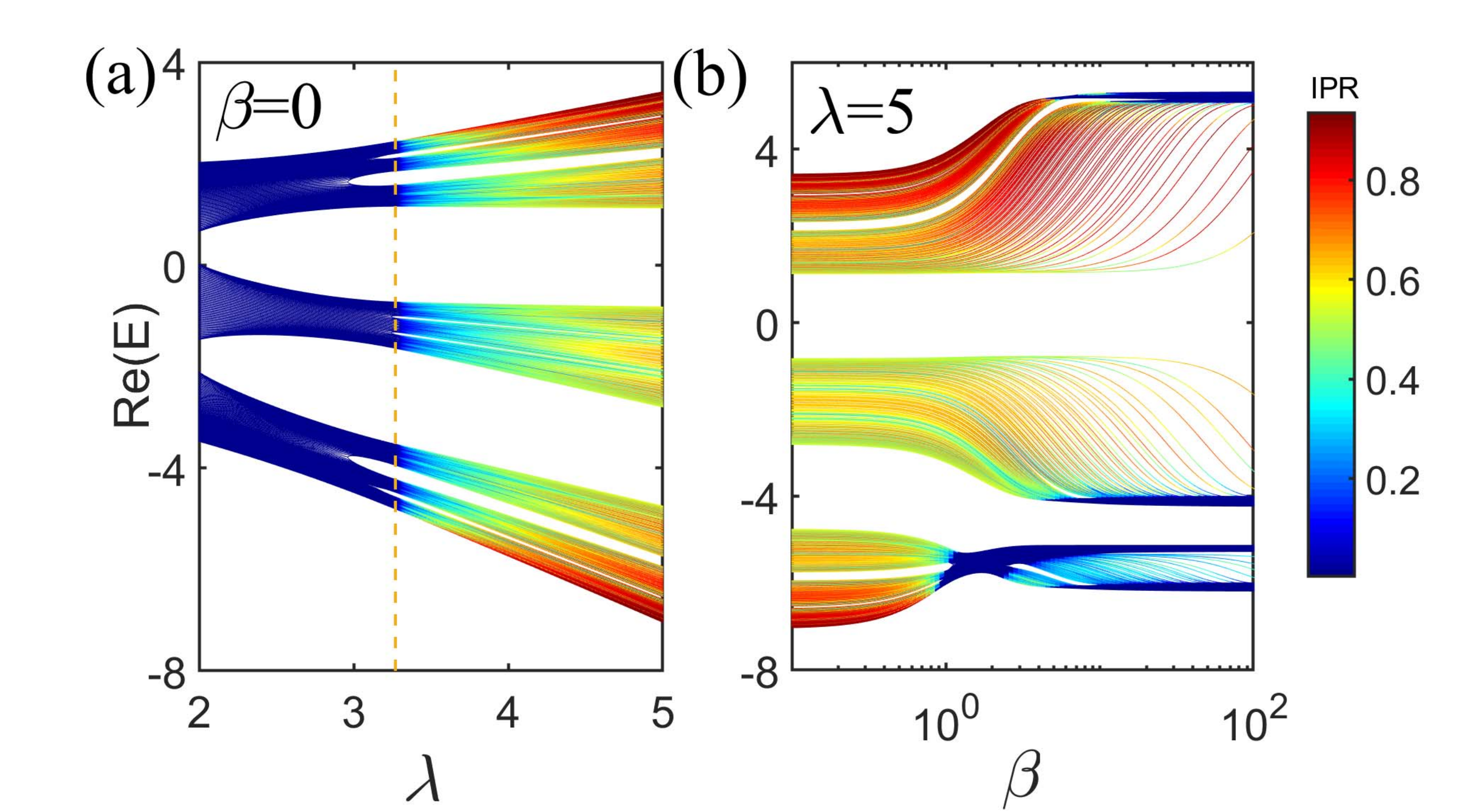}\\
  \caption{(a) ${\rm IPR}$ of all eigenstates of Eq.~(\ref{Eq:model}) as a function of the real part of the eigenenergy and $\lambda$ in the non-Hermitian AA limit.
  The dashed line is $\lambda_c=2e^{g}$.
  (b) ${\rm IPR}$ of all eigenstates of Eq.~(\ref{Eq:model}) as a function of the real part of the eigenenergy and $\beta$.
  $g=0.5$ are used in (a), and $g=0.5$ and $\lambda=5$ in (b).
  The lattice size is $L=610$, and the PBC is assumed.}
  \label{fig:energy-IPR}
\end{figure}

Besides ${\rm IPR}_g$, similar cascade structures can also be found in the excited states.
In the AA model limit, the localization transition appears simultaneously at $\lambda_c=2e^{g}$, as shown in Fig.~\ref{fig:energy-IPR} (a).
By continuously tuning $\beta$ toward the Fibonacci limit starting from the strongly localized AA model,
we find that the lowest set of eigenstates delocalized firstly at $\beta\sim 1$ and then localized again at $2<\beta<10$, as shown in Fig.~\ref{fig:energy-IPR} (b).
In addition, for the higher sets of excited eigenstates, the cascade behavior does not appear for the present range of $\beta$ as shown in Fig.~\ref{fig:energy-IPR} (b).

\begin{figure}
  \centering
  \includegraphics[width=2.5 in]{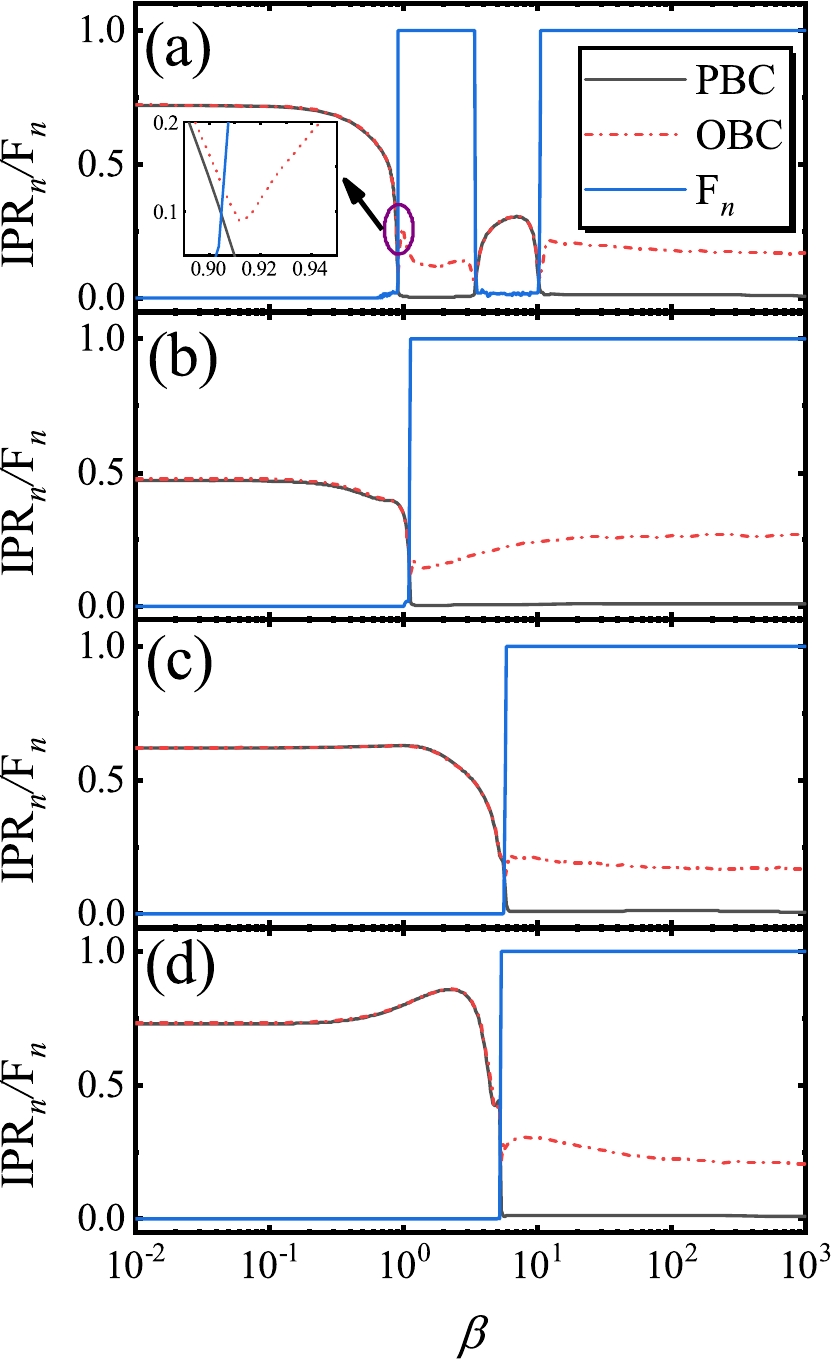}\\
  \caption{${\rm{IPR}}_n$ v.s. $\beta$ under PBC and OBC and ${\rm F}_{n}$ v.s. $\beta$ under PBC for $n$th excited states.
  Here, we use $\lambda=10$, $g=0.5$, $L=610$, and $n=$100, 200, 300, 500 for (a), (b), (c) and (d), respectively.
  ${\rm{IPR}}_n$ and ${\rm F}_{n}$ are averaged for 500 choices of $\phi$.
  The insert in (a) shows the minimum of ${\rm IPR}_n$ under OBC indicating the location of the critical state.}
  \label{IPRexcited}
\end{figure}

\begin{figure}
  \centering
  \includegraphics[width=2 in]{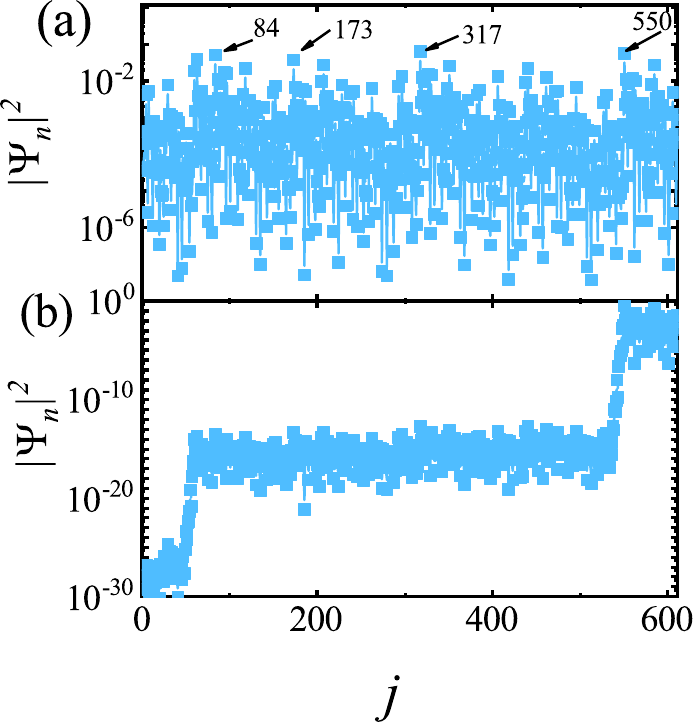}\\
  \caption{Typical spatial distribution of critical mode for excited state under PBC (top panel) and OBC (bottom panel).
    Here, we use $\beta=0.9311$, $L=610$, $g=0.5$, $\phi=0$ and $\lambda=5$, which locates between the first plateaux and the transition region
    in Fig.~\ref{IPRexcited} (a) (see the insert in Fig.~\ref{IPRexcited} (a)).}
  \label{fig:Dis_e100}
\end{figure}

It has been found that the localization transition is always accompanied with the the real-complex transition of the eigenenergies~\cite{zhai2020,Hamizaki2019}.
For the non-Hermitian IAAF model~\ref{Eq:model}, the ground state is real since the time-reversal symmetry is still preserved.
However, for the excited states, we find that the real-complex transition and the localization transition occur at the same point.
A function ${\rm{F}}_n$, measuring the value of the imaginary part of the energy, is defined as
%To measure the real-complex transition, ${\rm{F}}_n$ is defined as
\begin{eqnarray}
  {\rm{F}}_n=
   \left\{
             \begin{array}{lr}
             0,&{\rm{Im(E}}_n)=0,  \\
             \\
             1,&{\rm{Im(E}}_n)\neq0.
             \end{array}
\right.
\end{eqnarray}
In Fig.~\ref{IPRexcited}, curves of ${\rm{IPR}}_n$ versus $\beta$ under PBC and OBC, and ${\rm{F}}_n$ versus $\beta$ under PBC for different excited states are plotted. One finds that the cascade of the the delocalization transition is accompanied by the cascade of the real-complex transition.

For the excited critical states, the spatial distribution also have the self-similar structure similar to that of the ground state.
In Fig.~\ref{fig:Dis_e100}, the spatial distribution of the critical mode of the excited state under PBC and OBC are also plotted.
One finds that the relative distance between the peaks follows the Fibonacci sequence under PBC,
and it localizes on one side but not boundary under OBC.

\section{\label{secSum}Summary}
In this paper, we have studied the cascade of the delocalization transition and the critical behavior in a non-Hermitian IAAF model.
In the non-Hermitian AA limit, the system undergoes a delocalization transition at $\lambda_c=2e^{g}$.
At the critical point, the spatial distribution of ground state has a self-similar structure under the PBC. Under the OBC, the wave function of the critical mode in non-Hermitian AA limit localizes in one side
but not at boundary, which is different from the non-Hermitian skin effect.
By calculating the ${\rm{IPR}}_g$, we find that the ${\rm{IPR}}_g$ of the critical mode in non-Hermitian AA limit
scales as ${\rm{IPR}}_g\propto L^{\nu}$ with $\nu=-0.1189$. This demonstrated that the non-Hermitian AA model and the Hermitian AA model belong to different universality classes.
In the non-Hermitian Fibonacci limit, we find that the system is always in the extended phase for any finite $g$ and $\lambda$, since ${\rm{IPR}}_g$ scales as ${\rm{IPR}}_g\propto L^{-1}$.
By tuning $\beta$ continuously from the non-Hermitian AA limit into the non-Hermitian Fibonacci limit, the cascade of delocalization transition is found for both the ground and excited states,
but only a few plateaux appears.
These results demonstrate that the non-reciprocal hoppings can drastically change the cascade behavior in the IAAF model.
In addition, we have shown that the spatial distributions of the critical state between two plateau still have a self-similar structure. Moreover,
we have found that the real-complex transition also demonstrates cascade behavior, similar to the delocalization transition for the excited states.
Besides the non-reciprocal hopping, the non-Hermiticity can also be induced by the on-site gain/loss~\cite{Hamizaki2019,Longhi2019,Midya2018,Leykam2017,Novitsky2021}.
It should be interesting to study the delocalization transition and the critical behavior in the IAAF model with on-site gain/loss. We leave this for further studies.

\section*{Acknowledgments}
LJZ is supported by the Natural Science Foundation of Jiangsu Province (Grant No. BK20170309)
and National Natural science Foundation of China (Grant No. 11704161).
S.Y. is supported by the National Natural science Foundation of China (Grant No. 41030090).

\end{document}